\documentclass[conference,twocolumn]{IEEEtran}

\usepackage{amsmath, amsthm}

\usepackage{amssymb}
\usepackage{amsfonts}

\usepackage{graphicx}
\usepackage{subcaption}
\usepackage{float}

\usepackage[utf8]{inputenc}

\usepackage{hyperref}
\hypersetup{
    colorlinks,
    linkcolor={black!50!black},
    urlcolor={blue!80!black}
}
\usepackage[ruled,vlined]{algorithm2e}
\SetAlCapSkip{1em}
\SetKwInput{KwInput}{Input}
\SetKwInput{KwOutput}{Output}

\usepackage{booktabs}
\usepackage{multicol}
\usepackage{longtable}
\usepackage{tabto}
\usepackage{enumitem}

\usepackage{blindtext}
\usepackage{framed}
\usepackage{tcolorbox}
\usepackage[]{quoting}
\usepackage{array}
\usepackage[thinlines]{easytable}
\usepackage{ifthen}
\newboolean{showcomments}
\setboolean{showcomments}{true} % toggle to show or hide comments
\ifthenelse{\boolean{showcomments}}
  {}
  
\usepackage[normalem]{ulem}

\theoremstyle{definition}

\theoremstyle{remark}

\usepackage{tikz}
\usepackage{collcell}
\usepackage{rotating}
\usepackage{xspace}

\newcommand{\NAME}{{{\sc Fact Fortress}}\xspace}

\begin{document}

\title{[WIP] Deploying ZKP Frameworks with Real-World Data: Challenges and Proposed Solutions}

\author{\IEEEauthorblockN{Piergiuseppe Mallozzi}
\IEEEauthorblockA{\textit{UC Berkeley} \\
Berkeley, USA \\
mallozzi@berkeley.edu}\\\ \textit{------DRAFT------}}

\maketitle

\begin{abstract}
Zero-knowledge proof (ZKP) frameworks have the potential to revolutionize the handling of sensitive data in various domains. However, deploying ZKP frameworks with real-world data presents several challenges, including scalability, usability, and interoperability. In this project, we present \NAME~\footnote{\textbf{Fact Fortress Website and Demo}:\\\url{https://pierg.github.io/fact-fortress-web/}}, an end-to-end framework for designing and deploying zero-knowledge proofs of general statements. Our solution leverages proofs of data provenance and auditable data access policies to ensure the trustworthiness of how sensitive data is handled and provide assurance of the computations that have been performed on it. 

ZKP is mostly associated with blockchain technology, where it enhances transaction privacy and scalability through rollups, addressing the data inherent to the blockchain. Our approach focuses on safeguarding the privacy of data external to the blockchain, with the blockchain serving as publicly auditable infrastructure to verify the validity of ZK proofs and track how data access has been granted without revealing the data itself.

Additionally, our framework provides high-level abstractions that enable developers to express complex computations without worrying about the underlying arithmetic circuits and facilitates the deployment of on-chain verifiers. Although our approach demonstrated fair scalability for large datasets, there is still room for improvement, and further work is needed to enhance its scalability. By enabling on-chain verification of computation and data provenance without revealing any information about the data itself, our solution ensures the integrity of the computations on the data while preserving its privacy.
\end{abstract}


\maketitle

\section{Introduction}
Zero-knowledge proofs (ZKPs) are a powerful tool for enabling privacy-preserving computations and have a wide range of applications, including authentication, identity management, and data privacy. However, the current deployment of ZKP frameworks faces several challenges, such as the lack of trust in how data is handled, difficulty in constructing proofs, and scalability concerns.

In this project, we propose solutions to address these challenges and tackle three critical issues in ZKP-based applications. First, we address the problem of establishing trust in how data is handled by ensuring transparent and verifiable authenticity and provenance of data, implementing access controls based on data provider policies, and maintaining a public log for transparency. We propose a general statement verification process that includes proof of data provenance by design and an architecture of smart contracts to handle access control policies and the on-chain verification of proofs, providing accountability and transparency on the actions taken with the data.

Second, we address the problem of circuit generation, which involves creating an arithmetic circuit to formulate the desired function to be applied to the data. To simplify this process, our solution provides a user-friendly interface for generating circuits at scale. Our tool accepts a variety of data formats and a library of functions that can be chosen to analyze the data. The tool ultimately compiles down to circuits in the Noir language~\cite{noir}, a newly released Zero-Knowledge framework by Aztec. We evaluate the scalability of our approach and address how it handles data of growing size and circuits of growing complexity.

\begin{figure}[h]
    \centering
    \includegraphics[width=1\linewidth]{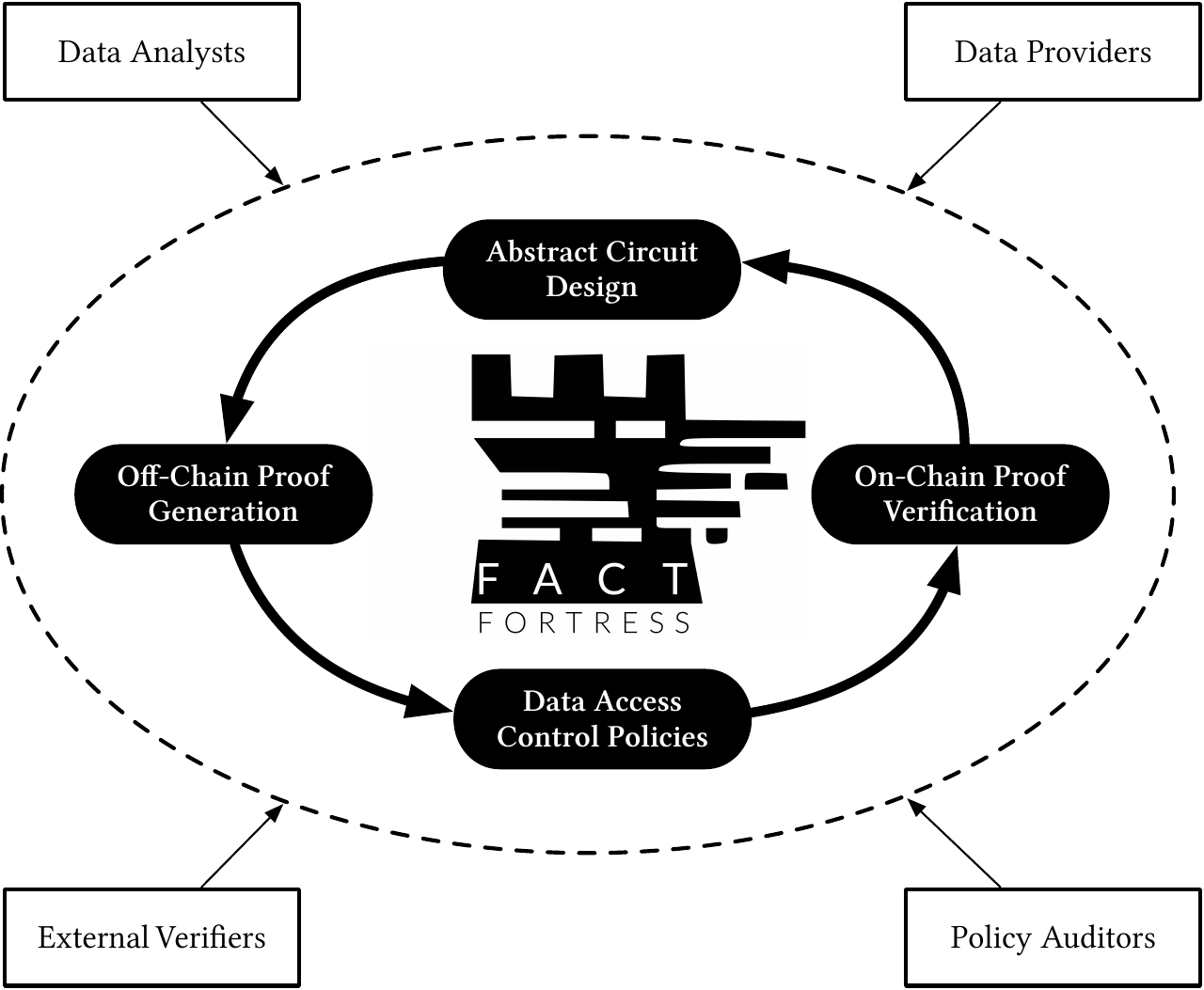}
    \caption{Overview of the proposed end-to-end solution.}
    \label{fig:end-to-end}
\end{figure}

Finally, our project provides a comprehensive solution that covers the entire process from circuit generation to proof generation, while facilitating collaboration among data analysts, data providers, external verifiers, and policy auditors, as illustrated in Figure~\ref{fig:end-to-end}.

We have implemented our approach in a platform named \NAME, a blockchain-integrated solution that utilizes ZKPs to enable efficient and trustworthy fact-checking with transparent and verifiable authenticity of data without compromising privacy. The platform is available open-source, and it includes a front-end and a separate tool for circuit compiler (code available\footnote{\textbf{Smart-Contracts Dapp}:\\\url{https://github.com/pierg/fact-fortress-dapp}\\\textbf{Front-end}:\\\url{https://github.com/pierg/fact-fortress-web}\\\textbf{Circuit Compiler Tool}:\\\url{https://github.com/pierg/fact-fortress-compiler}}).
\section{Background}

% In this section, we provide a brief overview of some fundamental cryptographic techniques: public-key encryption, digital signatures, and zero-knowledge proofs.

% \paragraph{\textbf{Public-key encryption}} is a cryptographic scheme that uses a pair of keys - one public and one private - to encrypt and decrypt messages. The public key is used for encryption and can be freely distributed, while the private key is used for decryption and must be kept secret. Examples of public-key encryption algorithms include RSA, Diffie-Hellman, and elliptic curve cryptography (ECC). This allows for \textit{secure communication} between parties without the need for a shared secret key.

% \paragraph{\textbf{Digital signature}} is a cryptographic technique that provides a way to verify the \textit{authenticity} and \textit{integrity} of digital messages or documents. 
% A digital signature is generated using a private key and can be verified using the corresponding public key. This allows the recipient of a message to verify that it was indeed sent by the claimed sender and that the message has not been tampered with.

\textbf{Zero-knowledge proof} is a cryptographic technique that allows one party (the prover) to prove to another party (the verifier) that they know a particular piece of information, without revealing that information itself. This can be especially useful in scenarios where sensitive data needs to be kept confidential to ensure the \textit{privacy} and \textit{security} of the data.

The process of zero-knowledge proof involves a prover and a verifier, who must agree on the function to be executed on the data. Let $f(w,x) = y$ be a function on the inputs $w$ and $x$ where $w$ (often called the \textit{witness}) is private and $x$ and $y$ are public. The prover must produce a proof $\pi$ that convinces the verifier that they know a secret input $w$ such that $f(w,x) = y$.

\begin{figure}[h]
    \centering
    \includegraphics[width=1\linewidth]{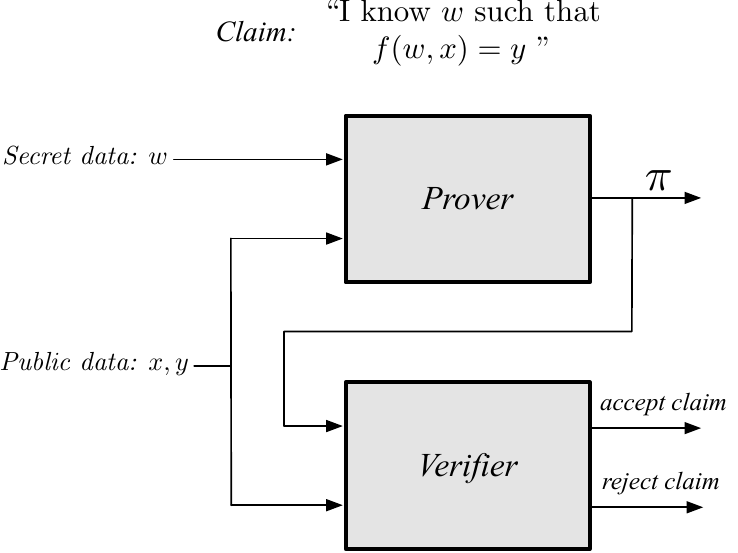}
    \caption{Generic Prover and Verifier Interaction}
    \label{fig:zkp-claim}
\end{figure}

Figure~\ref{fig:zkp-claim} illustrates the interaction between the prover and verifier to prove a generic claim $f(w,x) = y$ without revealing any information about $w$. The prover takes as input private data $w$ and public data $x$ and $y$, and generates a proof $\pi$. The verifier takes as input public data $x$, $y$, and $\pi$, and can either accept or reject the proof based on whether it is valid or not.

One specific implementation of zero-knowledge proof is called zk-SNARKS (Zero-Knowledge Succinct Non-Interactive Argument of Knowledge). It is a type of proof system where the prover produces a succinct and efficient proof of the correctness and the verifier is fast to verify the proof. The main steps to construct a zk-SNARK are:

\begin{enumerate}

\item \textbf{Arithmetic Circuit}: The function $f(w,x)=y$ needs to be represented as an arithmetic circuit consisting of multiplication and addition gates. This process is done by first converting $f$ into a R1CS (Rank-1 Constrain System), which is then transformed into a series of quadratic arithmetic programs (QAPs) using techniques such as Lagrange Interpolation and Fast Fourier Transform (FFT).
\item \textbf{Setup Procedure}: The setup procedure generates the proving and verifying keys $S_p$ and $S_v$, as well as the public parameters $pp$. The prover uses $S_p$ to generate proofs, the verifier uses $S_v$ to verify those proofs, and $pp$ defines the mathematical structure used to construct the circuit and is used by all parties. There are three types of setup procedures:
\begin{enumerate}
\item \textit{Trusted setup}, where a trusted party generates $(S_p, S_v, pp)$ for a specific circuit.
\item \textit{Trusted but updatable setup}, which is similar to the trusted setup but allows the trusted party to update parameters and keys for other circuits.
\item \textit{Transparent setup}, which generates parameters and keys using a publicly known deterministic algorithm, eliminating the need for a trusted party but may be more computationally expensive.
\end{enumerate}
\item \textbf{Prover}: The prover takes as input the private input $w$, the public data $x$ and $y$, the proving key $S_p$, and the parameters $pp$, and uses them to generate a proof $\pi$. This involves constructing a witness polynomial that satisfies the QAP and evaluating it at carefully chosen points, resulting in a proof consisting of two polynomials.
\item \textbf{Verifier}: The verifier takes as input the public data $x$ and $y$, the proof $\pi$, the verification key $S_v$, and the public parameters $pp$, and uses them to check the validity of the proof. This involves checking that the polynomials in the proof satisfy certain constraints, such as the QAP and the arithmetic relations of the circuit.
\item \textbf{Zero-Knowledge Property}: Finally, if the proof is valid, the verifier accepts the proof without learning anything about the private input $w$ or the computation of $f(w,x)=y$, except for the fact that the computation is correct.

\end{enumerate}

In summary, the setup procedure generates the proving and verifying keys and the public parameters, while the prover constructs a proof, and the verifier checks the validity of the proof. If the proof is valid, the zero-knowledge property guarantees that the verifier learns nothing about the private input.

\section{Challenges of Deploying Zero-Knowledge Proof Frameworks}

Several challenges hinder the adoption and deployment of ZKP frameworks in real-world applications. In this section, we discuss challenges of trust and adoption.

\subsection{Trust}

Trust is a critical challenge that must be addressed before zero-knowledge proof (ZKP) frameworks can be widely adopted in the real world. This is particularly important when working with sensitive data, where proving and verifying mechanisms should be transparent and the access to the sensitive data be regulated.

One challenge related to trust is ensuring the authenticity and provenance of the input data used by the prover and the verifier. If the input data used by the prover is compromised or tampered with, the resulting proof may be invalid, compromising the overall security of the system. On the other hand, the verifier needs to be confident that the public data used in the verification process is valid and has not been manipulated to produce a false result.

To address trust-related challenges, one approach is to include the input data as part of the trusted setup procedure. However, this requires trust in the entity that generated the public parameters. This technique can be used to deploy a trusted prover and verifier with fixed trusted data. However, it does not allow for the same prover and verifier to be used on other data, as the setup procedure would need to run again, and a new prover/verifier should be created.

Another approach is multi-party computation (MPC)~\cite{goldreich1998secure}, which allows multiple parties to jointly compute a function without revealing their inputs to each other. We can use MPC to collectively create the input data from several trusted entities. This technique can provide a more robust solution for generating trusted input data and can be used to deploy ZKP systems on a broader range of data without requiring a new trusted setup for each case, however it requires the coordination of multiple trusted entities.

% , particularly in ensuring that the input data originates from authenticated sources. Malicious actors could manipulate the input data, leading to a violation of the system's security and privacy guarantees. Techniques such as Zero-Knowledge Proofs of Knowledge (ZKPK) and distributed consensus mechanisms can help establish trust in the authenticity and integrity of input data.

% In summary, ZKP frameworks offer a promising solution to privacy and security challenges. However, the challenges of adoption, scalability, and trust must be carefully addressed to enable the deployment of ZKP in real-world applications. Research efforts are ongoing to overcome these challenges, and future developments in ZKP frameworks may lead to increased adoption and integration into various applications.

\subsection{Adoption}

The adoption of ZKP frameworks has been limited, in part, due to the complexity of expressing functions in terms of arithmetic circuits, which are a crucial component of many ZKP protocols. 
However, over the recent years, several tools and languages have been developed to facilitate the use of ZKP frameworks.

Hardware Description Languages (HDL) such as Circom~\cite{circom} provide a way to describe circuits in a low-level format that can be compiled to arithmetic circuits. Libraries such as Arkworks~\cite{arkworks} provide modular building blocks that can be combined to create custom ZKP systems.

Finally, there are several programming languages designed to make it easier to implement ZKP frameworks in real-world applications. For example, Zokrates~\cite{zokrates} is a popular open-source toolkit that allows developers to write programs in a high-level language and compile them into arithmetic circuits. Noir~\cite{noir} is a Rust-like language that provides tools for ZKP construction and verification. Leo~\cite{leo} and Cairo~\cite{cairo} are another programming languages that are designed to allow easy expression of complex circuits in a high-level format. 

Even with the availability of Domain-Specific Languages (DSLs), Programming Languages (PLs), and libraries, expressing complex functions in terms of arithmetic circuits remains a challenge for non-domain experts.

\section{\NAME}
In the following sections we provides an overview of our proposed solution named \NAME. In Section~\ref{sec:provenance} we address the problem of trust, providing general circuit design and smart-contract architecture that has data privacy and autheticity at its core. In Section~\ref{sec:circuits} we address the problem of adoption, we propose abstraction layers on top of existing frameworks that facilitate the circuit and data specification.
% In Section~\ref{sec:scalability} we present some scalability results of our approach.

\section{Trust by Design}
\label{sec:provenance}

\begin{figure}[h]
    \centering
    \includegraphics[width=1\linewidth]{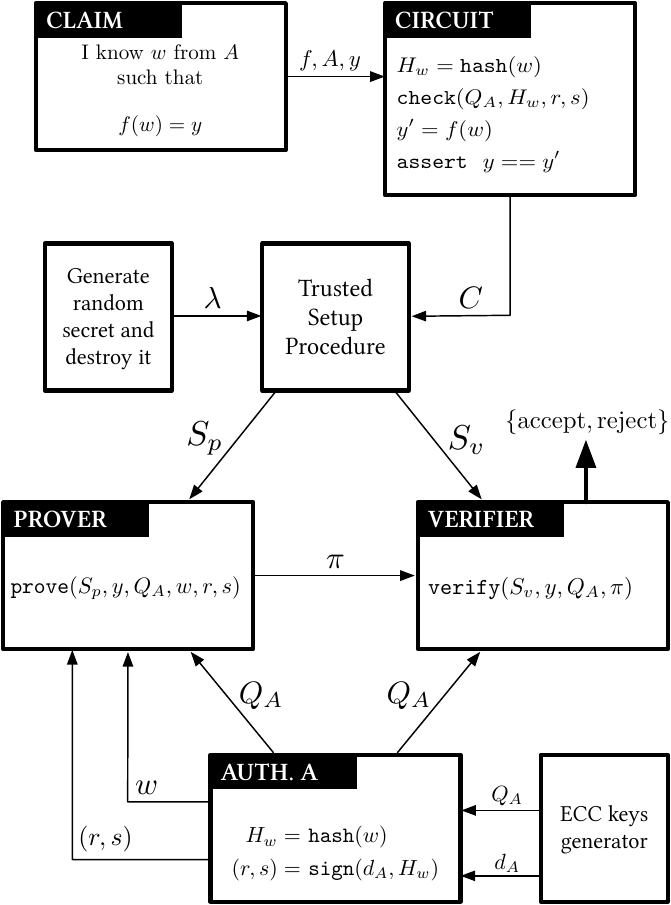}
    \caption{Statement Verification Process using ZKP}
    \label{fig:circuit-process}
\end{figure}

\begin{figure*}[h]
    \centering
    \includegraphics[width=1\textwidth]{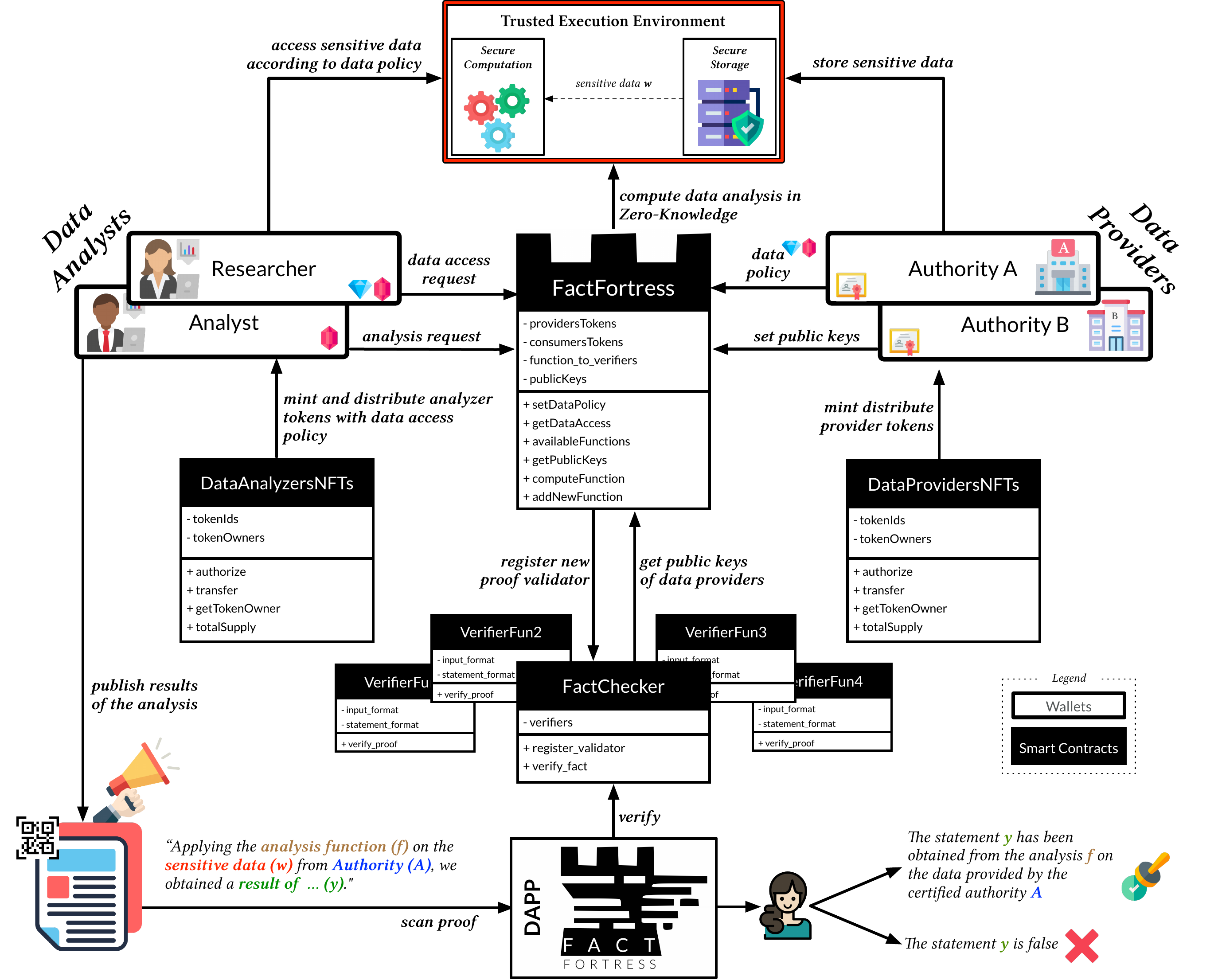}
    \caption{Smart Contracts Architecture on Blockchain}
    \label{fig:architecture}
\end{figure*}

In our approach, the trust of the data is embedded in the design of each circuit as well as in the overall framework.
Specifically, each circuit is designed to certify the proof of provenance of the data, ensuring that the input data is coming from authenticated sources. Moreover, the overall architecture framework is designed to facilitate the exchange of data among different parties in a regulated way implementing access control policies. Finally, we ensure accountability through a distributed ledger that stores all interactions among parties in a transparent and auditable manner.

By employing this our approach, the trustworthiness of the ZKP framework is enhanced, making it more suitable for handling sensitive data and real-world applications.

\subsection{Proof of Provenance}

Our approach involves embedding a ``\textit{proof of provenance}" alongside the proof of statement in each circuit. Figure~\ref{fig:circuit-process} provides an example of how we prove the truthfulness of a generic statement ``\textit{I know $w$ from $A$ such that $f(w)=y$}", without revealing any information about $w$. The proof is constructed in two steps:

Step 1: Proof of provenance, which proves that:
\begin{itemize}
\item $w$ originated from the authority $A$
\item $w$ has not been tampered with or altered in any way
\end{itemize}

Step 2: Proof of statement, which proves that:

\begin{itemize}
\item The result of $f(w)$ is equal to the claimed result $y$
\end{itemize}

In addition to the proofs provided in ZK by the circuit, our framework also ensures that `\textit{The function $f$ has been faithfully translated into an arithmetic circuit}' by providing a library of fixed functions that the prover can choose from (Section~\ref{sec:circuits}) and that `\textit{The proof $\pi$ was generated by the function $f$ claimed by the prover}' (Section~\ref{sec:architecture}).

\subsection{Data Access Policies}

To provide secure and regulated data access, our framework incorporates data-access policies. Each data provider can define their access policies, which include the type of data that can be accessed, the duration of access, and any restrictions on data usage.

Data analysts are granted granular data access permissions by attributing a Non-Fungible Token (NFT). Then, they can request access to the data at any time by providing proof of compliance with the access policies (\textit{i.e.}, the possession of the token, and the associated data access permissions). They are granted temporary access to the requested data if the proof is valid.

Our access control mechanism ensures that only authorized data analysts can access the data and that they can only use the data according to the specified policies. The NFTs also ensure that data analysts cannot access data that they are not authorized to use or extend their access beyond the specified duration.

By using smart contracts, we can guarantee the integrity and transparency of the data access policies and their enforcement, which is essential in sensitive domains such as government or finance.

\subsection{Accountability}

Our framework leverages the use of a transparent and publicly verifiable ledger, such as a blockchain, to ensure that data usage is logged and monitored in a transparent and immutable way. This creates an audit trail that tracks all data access and usage on the ledger, allowing data providers to monitor how their data is being used and to detect any unauthorized access or usage. By ensuring the integrity and transparency of data access and usage, our framework provides a robust and trustworthy solution for data sharing and analysis.

\subsection{Smart Contract-Based Architecture}
\label{sec:architecture}

To ensure transparency and accountability in data access, our \NAME framework incorporates smart contracts into its architecture. The overall architecture of the framework is depicted in Figure~\ref{fig:architecture}. By incorporating smart contracts into the architecture, we ensure that all parties involved in the data analysis adhere to the specified policies, and that the public results of the analysis can be trusted.

The framework allows certified \textit{data providers} to securely store their sensitive data and set data access policies on how the data must be handled. Data analysts can request access to the data based on these policies to perform an analysis and compute the ZKP locally. Alternatively, they can delegate the data analysis to \NAME, which returns the zero-knowledge proof of the computation and the result directly to them.

The types of analyses that can be performed are defined by a library of functions that can be computed on data of any form. For each function, we deploy a verifier on-chain, which anybody can use to validate a proof. When a proof is submitted for validation, \NAME dispatches it to the correct verifier, which checks its validity.

The overall process is as follows:
\begin{enumerate}
\item Our circuit compiler generates all the necessary data structures and circuits, ready to be proved in zero-knowledge, and produces a new on-chain verifier for each function in the library.
\item To \textbf{verify a proof}, the verifier (i.e. the \textit{Fact Checker} contract) receives the public keys of the authority and uses them as public inputs. The verifier ensures that: \begin{enumerate}
        \item The proof was generated by the function claimed by the data analyst (i.e., the prover).
        \item The data used to generate the proof has not been tampered with by the analyst and comes from a certified data provider.
        \item The claimed result is the correct result of the function applied to the data.
    \end{enumerate}
\item To \textbf{produce a proof}, the analysts can: \begin{enumerate}
    \item Securely access data via NFT-enabled access policies and produce the computation and proof themselves.
    \item Directly delegate \NAME to perform authorized functions in Zero-Knowledge in a trusted execution environment. The smart contract returns the result of the computation together with the proof.
\end{enumerate}
\item Analysts can confidently publish the results together with the proof. 
\item Anyone can read the results, download the proof, and publicly verify it on-chain.
\end{enumerate}

% \begin{figure*}[h]
%     \centering
%     \includegraphics[width=1\textwidth]{img/circuitsgen.png}
%     \caption{Circuits Generation Process}
%     \label{fig:circuitsgen}
% \end{figure*}

\section{Democratizing ZKP Circuits}
\label{sec:circuits}

\begin{figure}[h]
    \centering
    \includegraphics[width=1\linewidth]{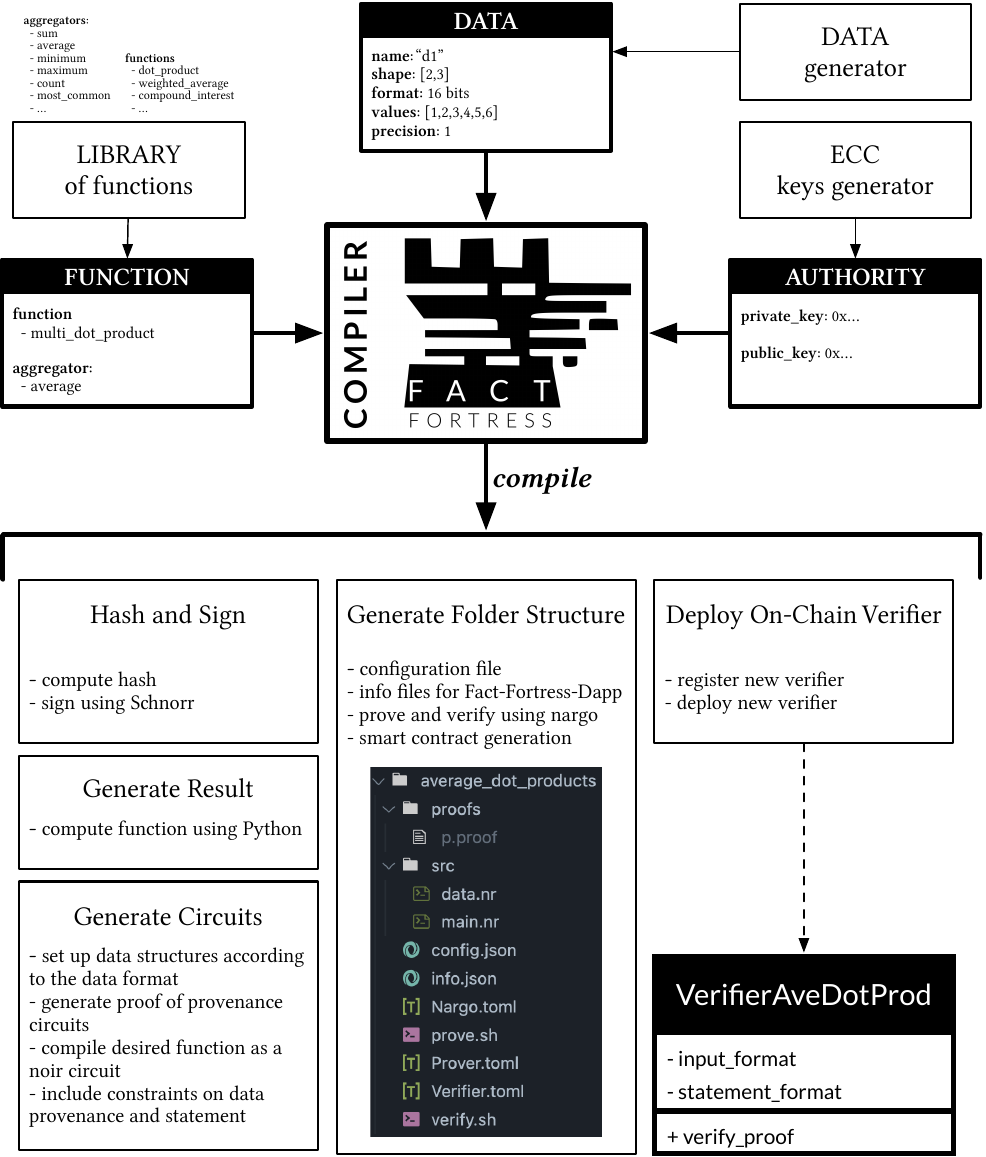}
    \caption{Circuit Compilation Process}
    \label{fig:compiler}
\end{figure}

To increase adoption of zero-knowledge proof (ZKP) frameworks and enable developers with limited experience to implement functions in real-world applications, more user-friendly interfaces and higher-level abstractions are needed to abstract away the low-level details of arithmetic circuits.

One approach to achieve this is through the development of tools for automatic circuit synthesis and by providing easy access for users to utilize them to verify their statements. In the following section, we describe how we have achieved this within our \NAME framework.

\subsection{\NAME Circuits Compiler}

We have implemented such a tool by building abstraction layers on top of the Noir framework~\cite{noir}. Our abstractions provide simple Python APIs that allow developers to express complex computations more easily, without requiring them to understand the underlying arithmetic circuits.

Our tool (code available\footnote{Configuration file: \url{https://github.com/pierg/fact-fortress-circuits/blob/main/circuits/average_dot_products/config.json}}) enables the automatic generation of arithmetic circuits from high-level abstractions, making it easier to implement zero-knowledge proof (ZKP) protocols in their applications.

Our library provides clear and abstract APIs that allow users to specify the data format, the function to be performed by the circuit, and the authority that provided the data. The library compiles down from Python API to JSON configuration file and ultimately parses the JSON and compiles the fully functioning circuit in Noir as shown in Figure~\ref{fig:compiler}.

In order to produce a circuit, our library takes as input:
\begin{itemize}
    \item \textit{data}: arrays of any size and shape, elements can be integers, string or double. Doubles are quantized by our library with the specified precision.
    \item \textit{authority}: private keys of the authority providing the data.
    \item \textit{function}: analysis that must be performed on the data. This function must be chosen by our library, e.g., one can do \textit{average of dot products}, \textit{weighted sums} etc. and compose multiple primitive functions together.
\end{itemize}

The compiler processes the data and generates all the necessary data structures and circuits, ready to be proved in zero-knowledge. Finally, we can deploy our function-specific on-chain verifier as a smart contract.

Specifically, our library performs the following operations:

\begin{enumerate}
    \item Compute the hash of data, sign the hash using an authority's private key using Schnorr signature protocol.
    \item Perform the chosen function on the data in Python to compare the result with the one executed by the circuit in Zero Knowledge.
    \item Generate a new comprehensive configuration file in JSON format to programmatically share and re-create the circuits.
    \item Generate the circuit! Given a configuration file, our library will generate a structured folder with all files needed to generate the proof in Zero-Knowledge using Noir. The circuit compiles and generates valid proofs right out of the box without having the user write anything in any domain-specific language. Specifically, our library can generate circuits on any data size and shape and can prove:
    \begin{itemize}
        \item Proof of Provenance: Compiles circuits that can compute the data hash using SHA256 and checks that the hash is valid and that the data comes from the authority using Schnorr Signature.
        \item Proof of Statement: Translates the chosen function into a valid circuit and checks that the function applied to the data results in the expected statement previously computed in Python.
    \end{itemize}
    \item Generate smart-contract verifier for the chosen function, ready to be deployed on-chain.
\end{enumerate}

Once the process has completed, the user can navigate to the generated folder and immediately prove and verify the compiled circuit, without any additional modification on the DSL source code.

% and run the following commands to prove and verify the circuit respectively:
% \begin{verbatim}
% # PROVE
% nargo prove p

% # VERIFY
% nargo verify p
% \end{verbatim}

\begin{figure}[h]
    \centering
    \includegraphics[width=1\linewidth]{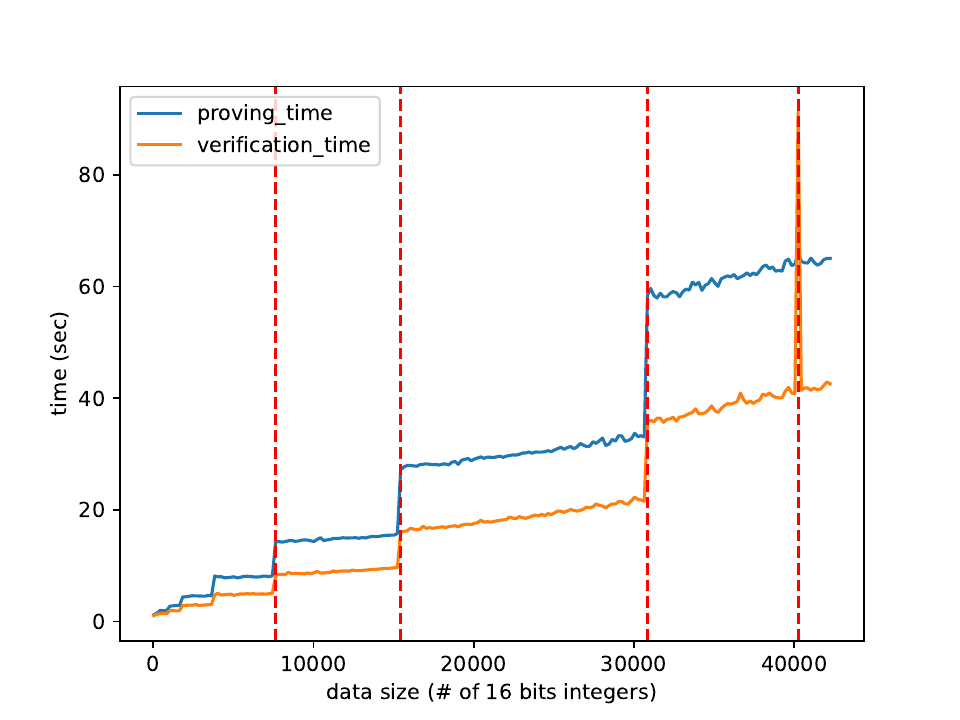}
    \caption{Scalability of prover and verifier. The x-axis represents the cumulative number of 16-bit integers in all data arrays, which corresponds to the size of the input data. The y-axis represents the proving and verification time in seconds.}
    \label{fig:plot-time}
\end{figure}

\subsection{Scalability}
\label{sec:scalability}
Scalability is a critical challenge for any ZKP framework, as the proving and verification times typically increase with the size of the input data and the number of arithmetic gates required by the function.

To address this challenge, we have conducted experiments on one of the functions from our library named \texttt{average\_dot\_products}, which takes as input a two-dimensional matrix and a vector. In this function, the dot-product between each row of the matrix and the vector is computed and then averaged over all rows.

We generated data for matrices and vectors of different sizes, where each element is a random integer of 16 bits. Figure~\ref{fig:plot-time} shows the proving and verification times for different sizes of data, as the number of operations increases with the data size.

We have observed that the proving and verification times increase with the size of the input data, as expected. Overall, the increase is sub-linear, indicating that our approach can scale well for large datasets. However, in certain regions (indicated by dotted lines), we have noticed a significant increase of the proving and verification times. To address this issue, we plan to conduct more experiments and explore additional optimization techniques to further improve the scalability of our approach in the future.

% As the size of the data increases, the proving and verification times grow sub-linearly. 
% Figure~\ref{fig:plot-operations} 

% This demonstrates that our approach is scalable and can handle large datasets and functions with a large number of arithmetic gates.

% We are continuously working to further optimize the performance of our framework and to explore new techniques to improve scalability.

% \input{tex/sections/implementation.tex}
\section{Conclusions}

ZKP is mostly associated with blockchain technology, where it enhances transaction privacy and scalability through rollups, addressing the data inherent to the blockchain. Our approach focuses on safeguarding the privacy of data external to the blockchain, with the blockchain serving as publicly auditable infrastructure to verify the validity of ZK proofs and track how data access has been granted without revealing the data itself. This provides a robust mechanism for preserving sensitive information privacy while leveraging blockchain technology's security and transparency.

In addition, our framework \NAME provides an efficient and user-friendly platform for designing and deploying zero-knowledge proofs of general statements. The high-level abstractions we provide enable developers to express complex computations without worrying about the underlying arithmetic circuits. We demonstrated that our approach scales well for large datasets, with a sub-linear increase in proving and verification times with the input data size. In the future, we plan to conduct more experiments and explore optimization techniques to improve the scalability of our approach further.

Our framework is open-source and available on GitHub. We believe that it can help democratize the use of zero-knowledge proofs and enable the development of more privacy-preserving applications.

\section*{Acknowledgments}
We would like to express our gratitude to Guillaume Lethuillier for his invaluable contributions to the validation of our proposed approach. His expertise in blockchain technology and insightful discussions have been instrumental in building and deploying the framework using Ethereum. 

We also express our gratitude to Brendan Hy for his contributions in building the front-end for the demo of this project. His expertise in web development and design greatly improved the user experience of our framework.

\bibliographystyle{IEEEtran}
\bibliography{main}

\end{document}